\documentclass{webofc}
\usepackage[varg]{txfonts}   % Web of Conferences font
\usepackage{hyperref}
\usepackage{url}
\usepackage{lineno}
\hypersetup{colorlinks=true,citecolor=blue,urlcolor=blue,linkcolor=blue}
\begin{document}
\title{Charmonium production in heavy-ion collisions}

\author{\firstname{Ingrid M.} \lastname{Lofnes}\inst{1}\fnsep\thanks{\email{ingrid.mckibben.lofnes@cern.ch}} for the ALICE Collaboration%\and
        %\firstname{Ingrid} \lastname{Lofnes}\inst{2}\fnsep\thanks{\email{Mail address for second
            % author if necessary}} \and
       % \firstname{Agnès} \lastname{Henri}\inst{3}\fnsep\thanks{\email{Mail address for last
          %   author if necessary}}
        % etc.
}

\institute{University of Bergen, Norway
%\and
%           the second here 
%\and
%           Last address
          }

\abstract{The early production of heavy quarks ($c\bar{c}$ and $b\bar{b}$) makes charmonia an ideal probe to study the evolution of the hot and dense medium produced in ultra-relativistic heavy-ion collisions, known as the quark--gluon plasma (QGP). 
At LHC energies, the well established suppression of charmonium yield from color screening in the QGP is counterbalanced by the recombination of uncorrelated charm-quark pairs, either throughout the QGP phase or solely at the phase boundary.
%At LHC energies, the recombination of uncorrelated charm-quark pairs has been found to significantly affect the measured charmonium observables, counteracting the well known suppression mechanism. 
Systematic measurements of charmonium ground and excited states are important to discriminate between the scenarios forseen by various theoretical models. 
In addition, studies of non-prompt charmonia, i.e., charmonia originating from the decay of beauty hadrons, give access to study the energy loss of beauty quarks in the QGP.
In these proceedings we present recently published Run 2 results by the ALICE Collaboration of prompt and non-prompt J/$\psi$ measured at midrapidity in Pb--Pb collisions at $\sqrt{s_{\rm NN}}$ = 5.02 TeV. In addition, the first preliminary Run 3 measurement of the $\psi$(2S)-to-J/$\psi$ ratio, measured at forward rapidity in Pb--Pb collisions at $\sqrt{s_{\rm NN}}$= 5.36 TeV, is shown. The results are compared to available theoretical model calculations. 
}
\maketitle
\section{Introduction}
\label{intro}
Heavy quarks (charm and beauty) are predominantly created during the initial hard partonic scattering and therefore experience the full evolution of heavy-ion collisions. They are therefore a natural probe when studying the hot and dense medium created in such collisions, known as the quark--gluon plasma (QGP).
Charmonium production, i.e., the creation of bound states consisting of charm-quark pairs ($\rm c\bar{c}$), is sensitive to the QGP. 
Color screening and medium induced dissociation may lead to a suppression of the measured charmonium yield in heavy-ion collisions with respect to binary-scaled proton-proton collisions. As the collision energy increases, the number of produced charm quarks will increase and the (re)combination of uncorrelated charm-quark pairs may lead to an enhancement of the measured yield. 

Experimentally, it is possible to separate the prompt and non-prompt charmonium production based on their production vertex. Here, prompt charmonium refers to either directly produced charmonium states or contributions from higher excited charmonium states, while non-prompt charmonium refers to states originating from the decay of beauty-hadrons. Measuring separately the prompt and non-prompt charmonium yields give access to different physics aspects of the QGP. The prompt charmonia are sensitive to charm thermalization in the medium and contributions from (re)generation and dissociation, while measurements of non-prompt charmonia give insight to the energy loss of beauty hadrons in the QGP.

The ALICE detector is the only LHC experiment which can detect inclusive charmonia down to vanishing transverse momentum ($p_{\rm T}$) in central heavy-ion collisions. The measurements are carried out in the central ($|y| < 0.9$) and forward rapidity ($2.5 < y < 4$) intervals, through the dielectron and dimuon channel, respectively. At midrapidity, the time projection chamber (TPC) is used for tracking and particle identification and the inner tracking system (ITS) is used for tracking and vertex reconstruction. In particular, the two innermost layers of the Run 2 ITS configuration consisted of silicon pixel detectors, enabling the disentanglement of prompt and non-prompt J/$\psi$ mesons, i.e., the charmonium lowest-lying vector state. During Run 2, the determination of centrality was provided by the V0 detectors consisting of two scintillator arrays covering the forward ($2.8 < \eta < 5.1$) and backward ($-3.7 < \eta < -1.7$) pseudorapidity regions. 
At forward rapidity, the muon chambers (MCH) are used for track reconstruction and the muon identifier (MID) for muon identification.

%At forward rapidity, the muon spectrometer is used for muon triggering and tracking. For Run 3 measurements, the newly installed muon forward tracker (MTF) allows for prompt and non-prompt charmonium separation also in the forward rapidity region.

% Detector setup run 2 and run 3

\section{Charmonium results: selected highlights}

\begin{figure}[h]
\centering
\includegraphics[width=5cm,clip]{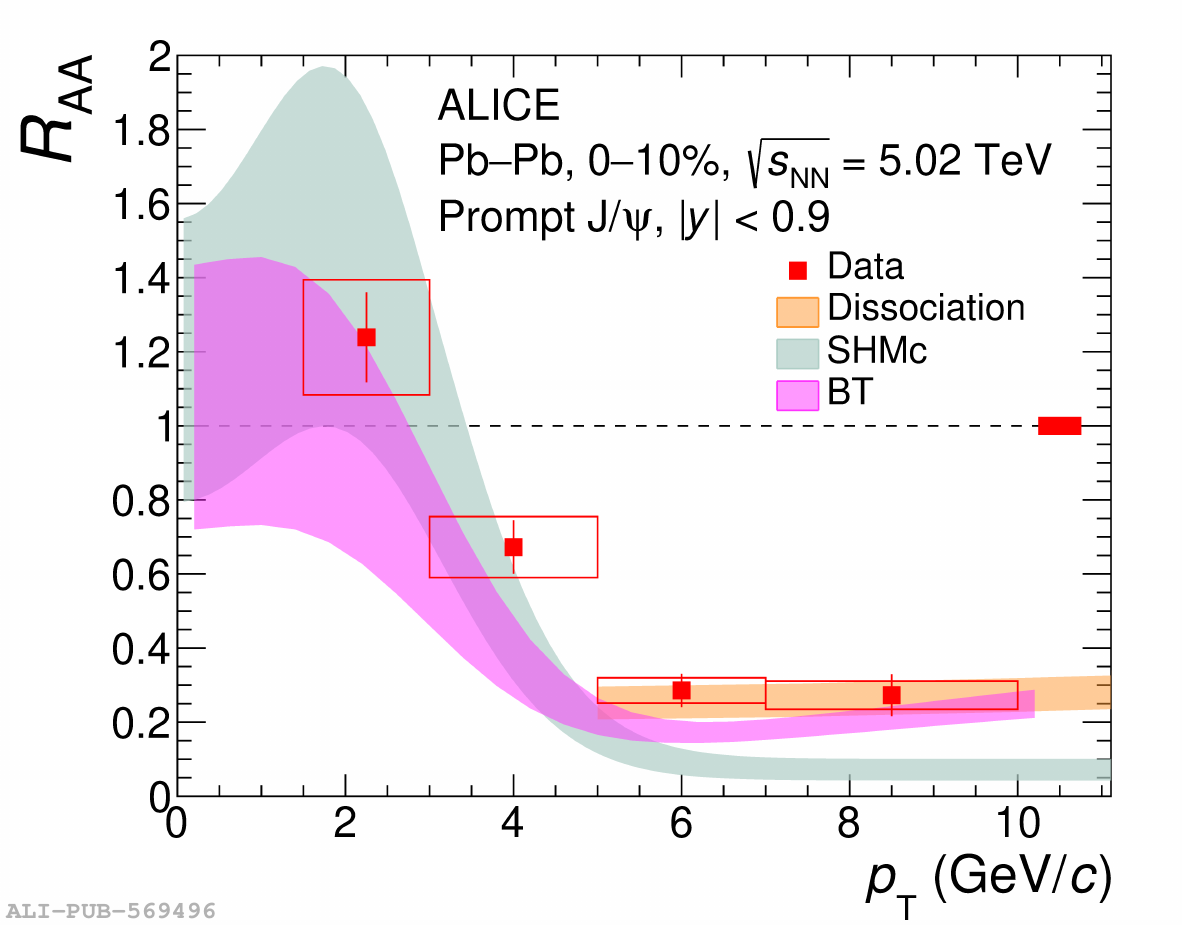}
\includegraphics[width=5cm,clip]{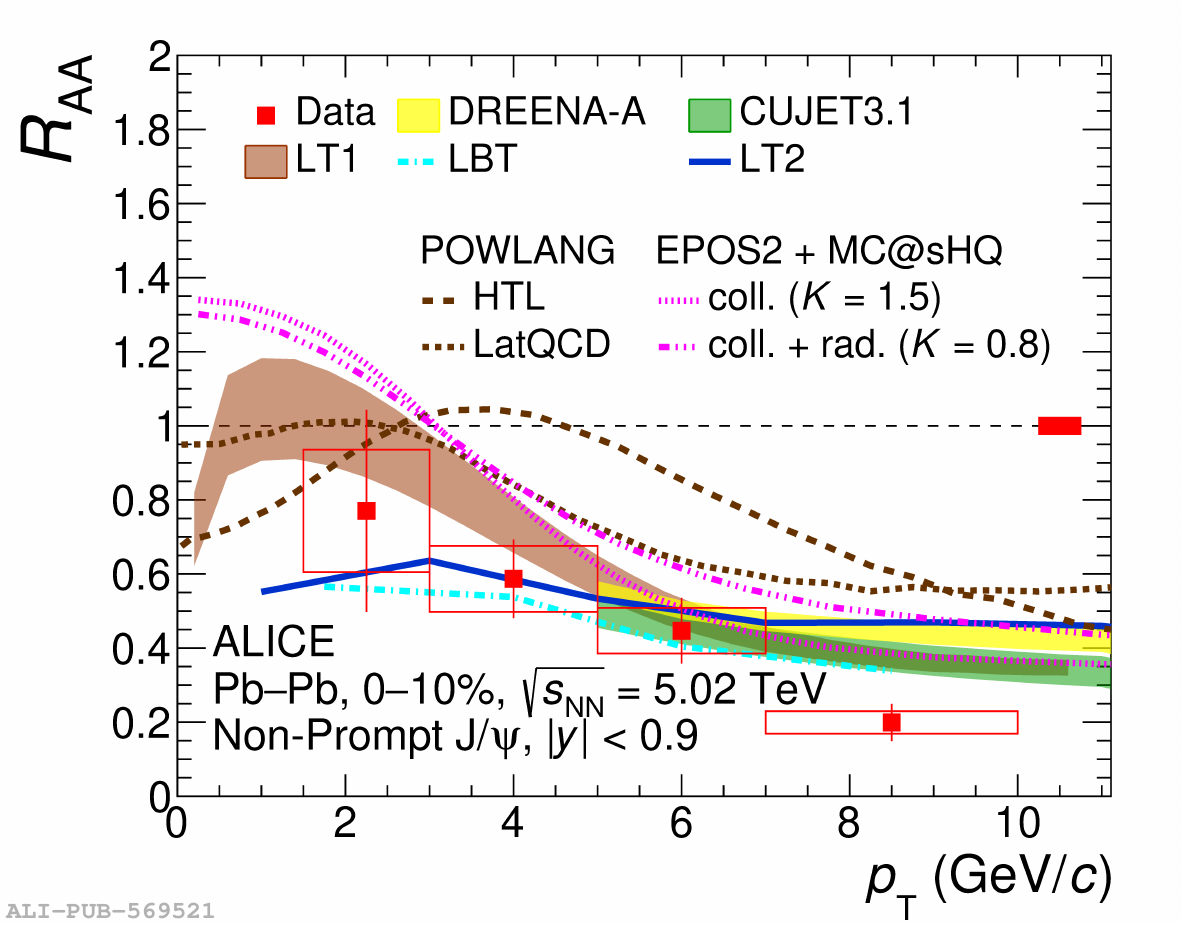}
\caption{$R_{\rm AA}$ as a function of $p_{\rm T}$ for prompt (left) and non-prompt (right) J/$\psi$ measured at midrapidity in Pb--Pb collisions at $\sqrt{s_{\rm NN}}$ = 5.02 TeV~\cite{ALICE:2023hou}. Theoretical model predictions for prompt~\cite{Zhou:2014kka,  Andronic:2019wva, Aronson:2017ymv} and non-prompt~\cite{Yang:2023rgb, Xing:2021xwc, Li:2021xbd, Nahrgang:2016lst, Zigic:2021rku, Shi:2018izg, Beraudo:2021ont} J/$\psi$ are also shown.}
\label{fig-1}       % Give a unique label
\end{figure}

%Discuss a bit run 2 prompt and non-prompt results
The nuclear modification factor ($R_{\rm AA}$) is defined as the ratio between the production yield in nuclear--nuclear collisions and the production cross section in proton--proton (pp) collisions scaled by the nuclear overlap function~\cite{ALICE:2018tvk}. Figure~\ref{fig-1} shows the Run 2 measurement of prompt and non-prompt J/$\psi$ $R_{\rm AA}$ in Pb--Pb collisions at $\sqrt{s_{\rm NN}}$ = 5.02 TeV~\cite{ALICE:2023hou} measured at midrapidity in the left and right panels, respectively.  
The $R_{\rm AA}$ is shown as a function of the transverse momentum ($p_{\rm T}$) for central collisions (0-10\%). The prompt J/$\psi$ $R_{\rm AA}$ shows a strong suppression for $p_{\rm T}>6$ GeV/$c$. Moving towards lower $p_{\rm T}$, the $R_{\rm AA}$ increases, reaching values above unity for $p_{\rm T}<3$ GeV/$c$. This is consistent with a larger contribution from the regeneration of uncorrelated charm quark pairs in the low-$p_{\rm T}$ region. The measurement is well reproduced by a Boltzmann transport model~\cite{Zhou:2014kka} in the full $p_{\rm T}$-range. In this calculation, Boltzmann-type rate equations are used to describe the dissociation and regeneration mechanisms. The statistical hadronization model (SHMc)~\cite{Andronic:2019wva} describes the measurement in the low-$p_{\rm T}$ region ($<6$ GeV/$c$), but  underpredicts the measurement at high $p_{\rm T}$. The SHMc extends the statistical hadronization to the charm sector, where uncorrelated charm-quark pairs are recombined solely at the phase boundary.
At high $p_{\rm T}$, the $R_{\rm AA}$ is well described by a dissociation model~\cite{Aronson:2017ymv} combining collisional charmonium dissociation and screening effects in the medium.

The non-prompt J/$\psi$ $R_{\rm AA}$ shows a strong suppression at high $p_{\rm T}$. In this region, models which include both collisional and radiative energy-loss processes describe the data within uncertainties~\cite{Yang:2023rgb, Xing:2021xwc, Li:2021xbd, Nahrgang:2016lst, Zigic:2021rku, Shi:2018izg}. The measurement is overpredicted by the POWLANG calculation~\cite{Beraudo:2021ont}, in particular when using hard thermal loops (HTL). The POWLANG calculation is the only model not including radiative energy-loss processes. At low $p_{T}$, the models describe the overall trend, but differ in the predicted magnitude of suppression. However, with the current precision it is not possible to discriminate between the different models.

\begin{figure}[h]
\centering
\includegraphics[width=4.4cm,clip]{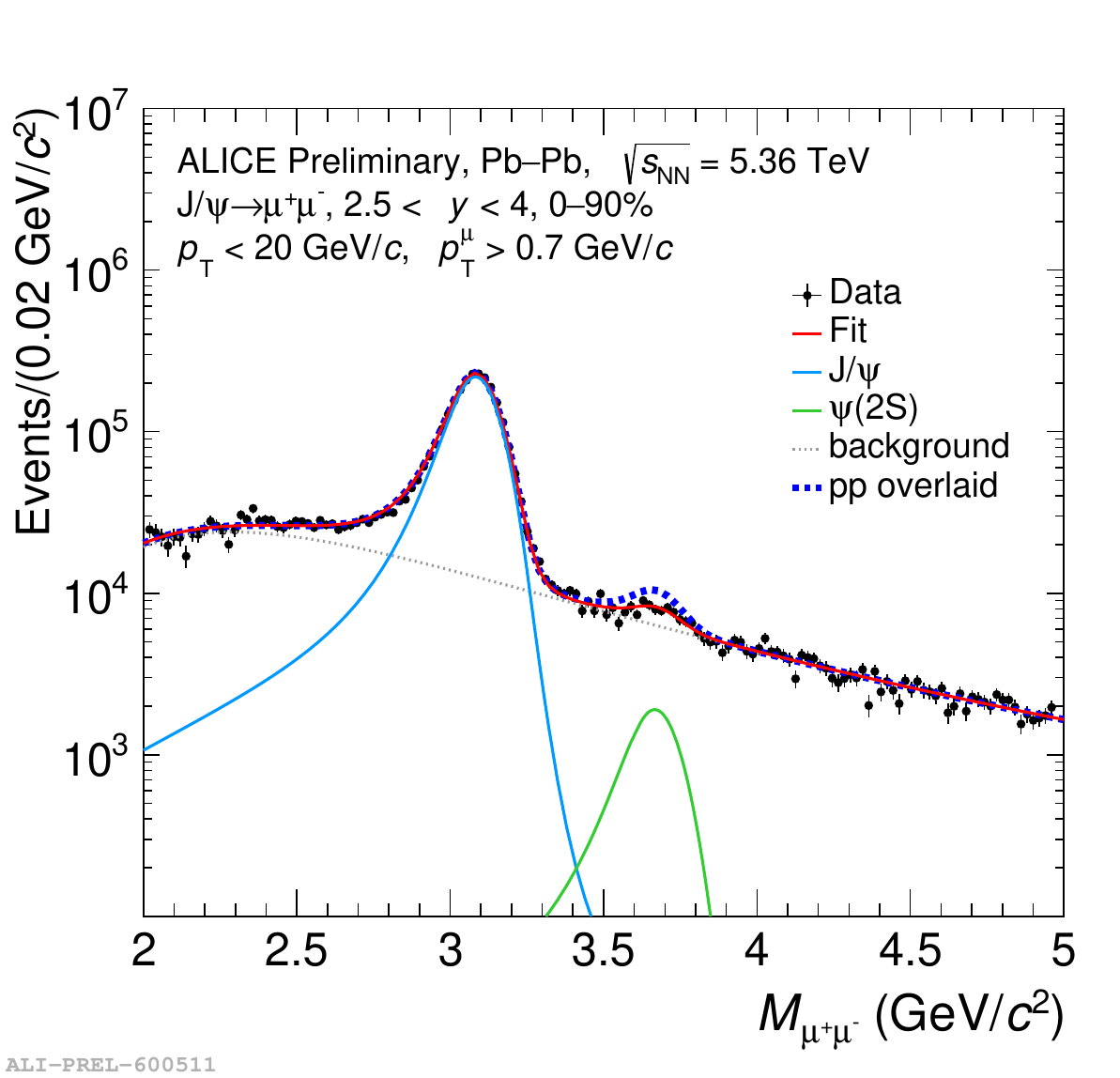}
\includegraphics[width=5.6cm,clip]{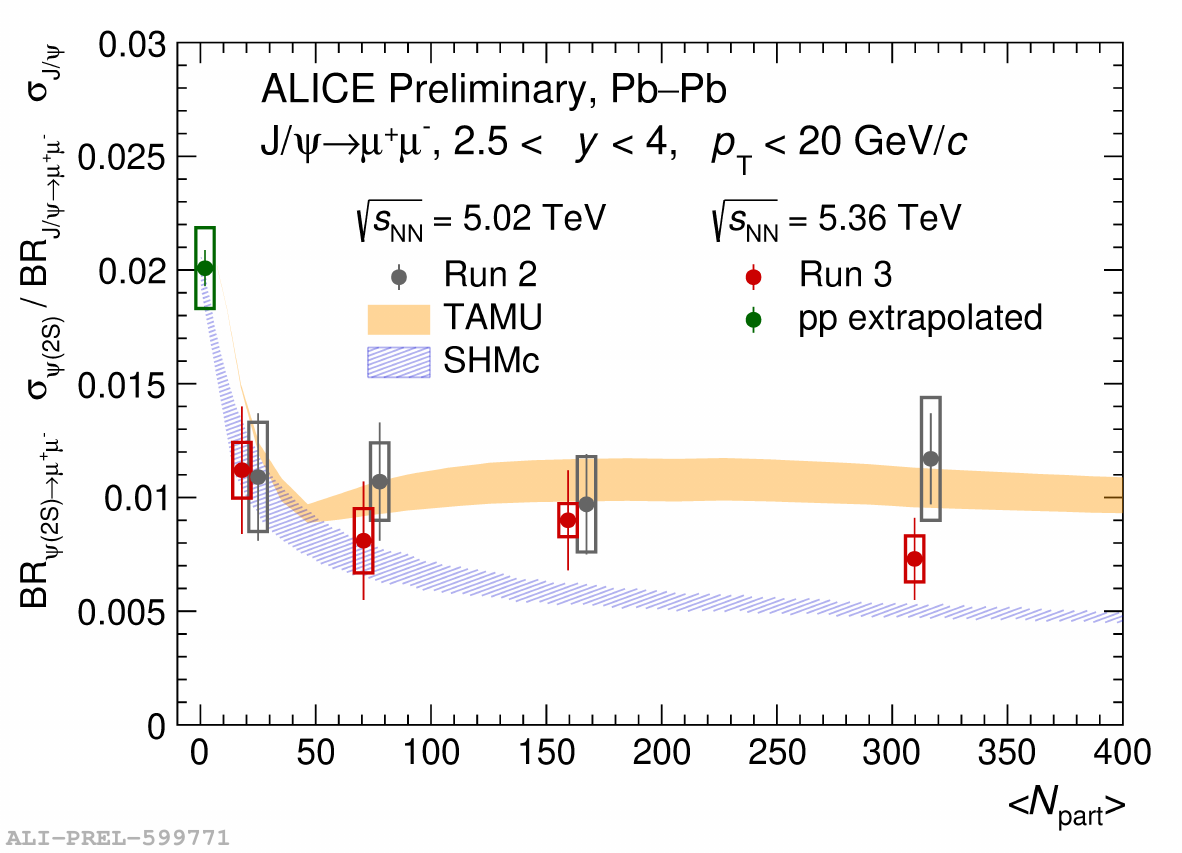}
\caption{Left: Invariant mass distribution of $\mu^{+}\mu^{-}$ pairs in Pb--Pb collisions at $\sqrt{s_{\rm NN}}$ = 5.36 TeV. The signal (blue and green) and background (gray) fit components are shown separately. In addition, the $\psi$(2S) component rescaled according to expectations without the QGP is shown (dashed blue). Right: $\psi$(2S)-to-J/$\psi$ ratio as a function of $\langle N_{\rm part} \rangle$ measured by ALICE at forward rapidity in Pb--Pb collisions at $\sqrt{s_{\rm NN}}$ = 5.02 TeV~\cite{ALICE:2022jeh} (green) and 5.36 TeV (red). The measurements are compared to TAMU~\cite{Du:2015wha} and SHMc~\cite{Andronic:2019wva} calculations. }
\label{fig-3}       % Give a unique label
\end{figure}

The preliminary Run 3 measurement of the J/$\psi$ and $\psi$(2S) in Pb--Pb collisions at $\sqrt{s_{\rm NN}}$= 5.36 TeV measured at forward rapidity is shown in Fig.~\ref{fig-3}. The left panel shows the invariant mass distribution of selected muon pairs after uncorrelated background subtraction, showing clear peaks around the J/$\psi$ and $\psi$(2S) mass above a residual background distribution. To visualize what the $\psi$(2S) peak would look like without medium-induced effects, an extrapolation of the pp ratio at the Pb--Pb collision energy of $\sqrt{s_{\rm NN}}$= 5.36 TeV is overlaid to the distribution to rescale the $\psi$(2S). The measured $\psi$(2S) is clearly suppressed with respect to the pp-scaled distribution. 

The right panel of Fig.~\ref{fig-3} shows the preliminary Run 3 measurement of the $\psi$(2S)-to-J/$\psi$ ratio in Pb--Pb collisions at $\sqrt{s_{\rm NN}}$= 5.36 TeV as a function of $\langle N_{\rm part} \rangle$. The ratio in Pb--Pb is suppressed by about a factor 2 with  respect to  pp collisions, and shows a flat behavior with centrality within current uncertainties. The preliminary Run 3 measurement is compatible to previous Run 2 results~\cite{ALICE:2022jeh}, but with higher precision. The improved precision may help in discriminating between different theoretical approaches.
The ratio is compared to a TAMU~\cite{Du:2015wha} and SHMc~\cite{Andronic:2019wva} calculation. The TAMU calculation is a transport model prediction, which includes color screening in the medium and recombination. In this model, a clear hierarchy in the suppression of charmonium states is employed. %The SHMc model predicts that the production of charmonium states is proportional to the total charm quark number. 

\section{Conclusions}
In these proceedings we investigate the production of charmonium states in heavy-ion collisions. The Run 2 measurement of prompt and non-prompt J/$\psi$ $R_{\rm AA}$ is studied in Pb--Pb collisions at $\sqrt{s_{\rm NN}}$ = 5.02 TeV. The prompt J/$\psi$ $R_{\rm AA}$ is consistent with contributions from regeneration and color screening, while the non-prompt J/$\psi$ $R_{\rm AA}$ is well reproduced by models implementing both collisional and radiative  energy-loss mechanisms in the QGP. 
The preliminary Run 3 measurement of the $\psi$(2S)-to-J/$\psi$ ratio in Pb--Pb collisions at $\sqrt{s_{\rm NN}}$ = 5.36 TeV shows a flat centrality dependence, and a factor 2 stronger suppression with respect to pp collisions.
With the newly installed muon forward tracker (MFT), separation of prompt and non-prompt charmonium will also become possible in the forward rapidity region, allowing for several interesting measurements in the near future. 

%
% BibTeX or Biber users please use (the style is already called in the class, ensure that the "woc.bst" style is in your local directory)
 \bibliography{bibliography.bib} % Replace "your_bib_file" with the actual name of your .bib file

\end{document}